\documentclass[12pt,preprint]{aastex}
\usepackage{graphicx}
\usepackage{url}

\begin{document}


\title{Consequences of Energetic Magnetar-Like Outbursts of Nearby Neutron Stars: $^{14}$C Events and the Cosmic Electron Spectrum}

\author{F. Y. Wang$^{1,2}$, Xinyu Li$^{3,4,5}$, D. O. Chernyshov$^{6}$, C. Y. Hui$^{7}$, G. Q. Zhang$^1$, K. S. Cheng$^{8}$}
\affil{
$^1$ School of Astronomy and Space Science, Nanjing University, Nanjing 210093, China fayinwang@nju.edu.cn\\
$^2$ Key Laboratory of Modern Astronomy and Astrophysics (Nanjing University), Ministry of Education, Nanjing 210093, China\\
$^3$ Department of Physics, Columbia University, New York, NY
10027 \\
$^4$Canadian Institute for Theoretical Astrophysics, 60 St George St, Toronto, ON M5R 2M8 \\
$^5$Perimeter Institute for Theoretical Physics, 31 Caroline Street North, Waterloo, Ontario, Canada, N2L 2Y5 \\
$^6$ I.E. Tamm Theoretical Physics Division of P.N. Lebedev
Institute of Physics, 119991 Moscow, Russia\\
$^7$ Department of Astronomy and Space Science Chungnam National
University, Daejeon 34134, Republic of Korea\\
$^8$ Department of Physics, University of Hong Kong, Pokfulam Road,
Hong Kong, China hrspksc@hku.hk}

\begin{abstract}

Four significant events of rapid $^{14}$C increase have taken place
within the past several thousand years. The physical origin of these
rapid increases is still a mystery but must be associated with
extremely energetic cosmic processes. Pulsars are highly magnetized
neutron stars that emit a beam of electromagnetic radiations. Any
sudden release of the energy stored in the magnetic multipole field
will trigger outbursts similar to the giant flares of magnetars.
Here we show that the relativistic outflow from the outbursts of a
nearby pulsar interacting with the interstellar medium generates a
shock, which accelerates electrons to trillions of electron volts.
The high-energy photons from synchrotron emission of the shock
interact with Earth's atmosphere, producing the cosmogenic nuclide
$^{14}$C, which can cause the rapid $^{14}$C increases discovered in
tree rings. These same relativistic electrons can account for a
significant fraction of the cosmic electron spectrum in the trillion
electron volts energy range, as observed by space-borne satellites.
Since these outburst events can significantly affect our
environment, monitoring nearby pulsars for such outbursts may be
important in the future.
\end{abstract}


\section{Introduction}
Observations have revealed the existence of strong magnetic fields
on neutron stars, with dipole component ranging from
$10^{11}-10^{13}$~G for pulsars and $10^{14}-10^{15}$~G for
magnetars. Neutron stars might also have multipole components much
larger than its dipole field. The strong multipole components can be
a consequence of dynamos inside the young fast spinning neutron star
\citep{Thompson93} and they are buried under the surface by the mass
accretion \citep{Romani90,Cumming01}, especially the prompt
accretion of supernova fallback material
\citep{Chevalier89,Geppert99}. Magnetars are a special type of
neutron stars showing violent transient radiative phenomena,
including giant flares and soft gamma-ray bursts with luminosity up
to $10^{47}$~erg per second \citep{Kaspi17}. These outbursts are
characterized by their X-ray luminosity exceeding the spin-down
power of the neutron star, and it is believed they are powered by
the high magnetic energy stored inside the star \citep{Thompson95}.
Such bursts and outbursts are not limited to magnetars. Recent
observations have found that pulsars with high magnetic field also
exhibit magnetar-like activities, e.g. PSR J1846-0285
\citep{Gavriil08} and PSR 1119-6217 \citep{Gogus16,Archibald17}.
These events can be powered by the high magnetic energy of the
multipole components buried inside the neutron star.

The cosmogenic nuclide $^{14}$C is produced in atmosphere either by
high energy particles (e.g. protons) or $\gamma$-ray photons from
high-energy phenomena. Through the global carbon cycle, some of
$^{14}$CO$_2$ produced in the atmosphere can be retained in annual
tree rings \citep{Damon95,Usoskin06}. Recently, four events of rapid
increase of the $^{14}$C content occurred in AD 994, AD 775, BC 660
and BC 3371 were found using tree rings
\citep{Miyake12,Miyake13,Wang17,Park17,Buntgen18}. However, the
physical origin of these events is mysterious. Several models have
been proposed, such as gamma-ray emissions from nearby supernovae
\citep{Miyake12} or gamma-ray bursts (GRBs)
\citep{Hambaryan13,Pavlov13}, and large solar proton events (SPEs)
\citep{Miyake12,Melott12,Usoskin13,Wang17}. Some recent works
\citep{Mekhaldi15,Buntgen18} argued that large solar proton events
caused the $^{14}$C rapid increase events. However, there are
several problems with the interpretation of these $^{14}$C events as
solar proton events \citep{Neuhauser14}. First, there are no
definite historic records of strong aurorae or sunspots around AD
775 and AD 994 \citep{Stephenson14,Chai15}. Second, if the SPE
fluence of AD 775 increase, which is only a few times larger than
that of the Carrington event in 1859 \citep{Usoskin13}, would have
been solar origin, there must also be a rapid $^{14}$C increase by
Carrington event, which is not the case
\citep{Miyake12,Neuhauser14}. Third, Cliver et al. (2014) compared
energetics and spectrum of the 1956 solar proton event with AD 775
case, and found the inferred solar fluence ($>$30 MeV) value for
this event is inconsistent with the occurrence probability
distribution for solar proton events \citep{Usoskin17}, which casts
strong doubts on the solar interpretation for AD 775 event
\citep{Cliver14}. Fourth, whether the Sun can produce such large
proton events is still debated \citep{Schrijver12}. Recent study
shows that the occurrence frequency of superflares on solar-type
stars is once in about three thousands years, which is too low
\citep{Notsu19}. Fifth, charged particles from solar proton events
are affected by the geomagnetic field, and will produce $^{14}$C
symmetrically in both hemispheres. However, the latest measurements
of B\"{u}ntgenet al. (2018) showed that the $^{14}$C increase starts
at different times for the AD 775 event in the southern and northern
hemispheres. The short-duration outburst would shine on only one
side of the Earth, which would affect the time and intensity of
$^{14}$C increase at different places on Earth \citep{Guttler15}.

The corresponding radiations caused $^{14}$C rapid increase not only
can deplete ozone, but also damage communication, electronic, power
systems on ground and electronic equipment in satellites. Therefore,
study the physical origin of $^{14}$C events is crucial. Here, we
propose that the high-energy emissions from pulsar outbursts can
cause rapid $^{14}$C increase. High energy electrons are very
important probe of nearby cosmic ray sources. Recent discoveries of
an excess of electrons stimulated a number of works to discuss their
possible origins. Outbursts of pulsars, which also generate high
energy electrons, may be responsible for the irregularities in the
spectrum of the electrons observed by various space-borne
experiments such as AMS-02 \citep{ams}, Fermi-LAT \citep{fermi17},
DAMPE \citep{dampe} and CALET \citep{adr2018}.

In Section 2, we describe the physical Model of pulsar outbursts.
Section 3 presents the gamma-ray emissions from outbursts and
$^{14}$C events. Section 4 discusses the cosmic electron spectrum
contributed by the outbursts. We search the possible neutron stars
in our neighborhood in Section 5. Conclusions are given in Section
6. Throughout the paper, we adopt the convention $Q_a = Q/10^a$
using cgs units.

\section{The physical Model of magnetar outbursts}

In a pulsar with dipole field $B=10^{13}$~G, a magnetic multipole
can be buried in the core with much higher magnetic field. Assuming
the magnetic multipole has magnetic field strength $B_m=10^{16}$~G
with length scale $L=10^5$~cm, the magnetic energy associated with
this magnetic multipole is
\begin{equation}
E_m \sim \delta B^2 L^3 \sim 10^{47} B_{m,16}^2 L_5^3 ~\rm erg.
\end{equation}
Such magnetic energy is
expelled from the core through ambipolar diffusion on the time scale \citep{Beloborodov16}
\begin{equation}
t_{\rm amb} = 30 L_5^{2/5} B_{m,16}^{2/5}/B_{13}^{8/5}
(\rho/\rho_{\rm nuc})^{6/5} ~\rm Myr .
\end{equation}
The dissipation of core magnetic energy can induce sudden global changes
of magnetospheric structure which powers the
outbursts \citep{Thompson95}. The origin of the sudden magnetospheric
changes can be a result of crustal failure at the core-crust
boundary when the high magnetic field escaping from the core breaks
the crust \citep{Thompson95}. Another possible mechanism to trigger
the outburst is a slow buildup of magnetic energy in the
magnetoshpere followed by a sudden
release \citep{Lyutikov03,Gill10,Parfrey13}. When the high magnetic
field propagates outward in the crust, its high magnetic stress can
break the crust and initiate plastic
failure \citep{Beloborodov14,Li16}. The plastic failure causes surface
displacement magnetic footpoint and lead to twisting of the surface
magnetic field. If the twisting angle exceeds a critical value, the
magnetosphere becomes unstable and produces a flare \citep{Parfrey13}.
In that case, the magnetosphere undergoes tearing instability and
magnetic reconnection takes place with the launching of Alfv\'{e}n
waves and the ejection of plasmoids similar to the solar coronal
mass ejection. The magnetic reconnection occurs within a very short
time $\sim 100 R/c \sim 10^{-2}$~s \citep{Uzdensky11}, where $R$ is
the radius of neutron star.

Thompson \& Duncan (2001) proposed that the extraordinarily high
peak luminosity, and rapid variability of the initial spike emission
from giant flares of magnetars imply that the spike emission must
originate from a relativistically expanding fireball with an initial
Lorentz factor of at least several tens, in order to avoid the
pair-production problem. The observed light curve of the spikes of
SGR 1806-20 \citep{Hurley05} is well explained by emission from
relativistically expanding fireballs \citep{Yamazaki05}. Meanwhile,
the radio afterglow also can be fitted by a relativistic outflow
with a Lorentz factor of a few tens \citep{Wang05}. Therefore,
similar to giant flares of SGRs \citep{Hurley05}, we assume a
relativistic outflow with Lorentz factor $\Gamma_0\sim 10$ is
launched during the outburst of pulsars. The ejecta propagates into
the interstellar medium (ISM) and builds up a relativistic forward
shock and a reverse shock structure, including shock-accelerated
electrons with associated synchrotron emission. Synchrotron
radiation of the shocks can provide high-energy photons for rapid
$^{14}$C increase and the accelerated electrons can responsible for
the irregularities in the spectrum of cosmic electrons.

\section{Gamma-ray emission and $^{14}$C events}
We first calculate the high-energy emissions from the outbursts of
pulsars. From theory and numerical
simulations \citep{Blandford87,Sironi13}, Fermi acceleration of
charged particles in collisionless shocks leads to power-law
distributions of energetic particles, described by
$dN/d\gamma_e\propto \gamma_e^{-p}$, where $\gamma_e$ is the
electron Lorentz factor and $p=2.1-2.2$. In relativistic
collisionless shocks case, the Bohm approximation is not valid for
maximum energy electrons \citep{Kirk10,Lemoine13}. The residence time
scale downstream is given by $t_{r}=N t_{\lambda}=(R_L/\lambda)^2
(\lambda/c)=(R_L/\lambda)(R_L/c)$, where $R_L$ is Larmor radius,
$N=(R_L/\lambda)^2$ is the scattering times that the electrons
experience before returning to the upstream, $\lambda$ is the size
of coherence cell and $t_\lambda$ is the time spent in crossing the
coherence cell. The Lorentz factor of the maximal energy electrons
can be derived by equating the synchrotron cooling time
$t_{syn}=6\pi m_\emph{e} c/(\sigma_{\rm T}\gamma_e B^2)$ to the
residence time $t_{r}$, which reads \citep{Kirk10}
\begin{equation}
\gamma_{e,max}=(6\pi \lambda e^2/\sigma_{\rm T} m_e
c^2)^{1/3}=1.2\times10^7 n_{0}^{-1/6}\lambda_1^{1/3},
\label{gamma_emax}
\end{equation}
where $\lambda\equiv 10 \lambda_1 c/\omega_{pi}$ with
plasma frequency $\omega_{pi}$ and $n$ is the number density of the
ISM.

When the relativistic outflow encounters the ISM, the external shock
is generated, including forward and reverse shocks \citep{Sari98}. The
standard afterglow model of gamma-ray bursts \citep{Sari98} is used.
We use the standard afterglow model of gamma-ray
bursts \citep{Meszaros02,Wang15,Zhang19}. $\epsilon_e$ and
$\epsilon_B$ are the shock energy equipartition parameters for the
shock accelerated electrons and the magnetic fields, respectively.
For the forward shock emission, the cooling frequency $\nu_{\rm c}$,
the typical synchrotron frequency $\nu_{\rm m}$ and the maximum
spectral flux $F_{\rm \nu,max}$ are \citep{Sari98}
\begin{equation}
\nu_{\rm c}=1.0\times 10^{20}{\rm Hz}~E_{\rm
iso,47}^{-1/2}\epsilon_{\rm B,-2}^{-3/2}n_0^{-1}{t}_{\rm
obs,1}^{-1/2}(1+Y)^{-2}, \label{nu_c}
\end{equation}
\begin{equation}
\nu_{\rm m}=8.4\times 10^{15}{\rm Hz}~C_{\rm p}^2E_{\rm
iso,47}^{1\over 2}\epsilon_{\rm B,-2}^{1\over 2}\epsilon_{\rm
e,-1}^2{t}_{\rm obs,1}^{-{3\over 2}}, \label{nu_m}
\end{equation}
\begin{equation}
F_{\rm \nu,max}=4.9\times10^7 {\rm Jy}~E_{\rm iso,47}\epsilon_{\rm
B,-2}^{1/2}n_0^{1/2}({d \over150~{\rm pc}})^{-2}, \label{Fmax}
\end{equation}
where $C_p=3(p-2)/(p-1)$ and $Y$ is the inverse Compton parameter,
which can be calculated from $Y\simeq [-1+\sqrt{1+4x\epsilon_{\rm
e}/\epsilon_{\rm B}}]/2$ \citep{Sari01}, where $x={\rm
min}\{1,2.67(\gamma_{\rm m}/\gamma_{\rm c})^{\rm (p-2)}\}$ is the
radiation coefficient of the shocked electrons, and $\gamma_{\rm c}$
is the electron cooling Lorentz factor \citep{Sari98}
\begin{equation}
\gamma_{\rm c}\approx {7.7\times 10^{8}\over (1+Y)}{1\over \Gamma
B'^2 t_{\rm obs}}.
\end{equation}
We estimate the observed flux at 20 MeV as
\begin{eqnarray}
f_{\nu_{\rm obs}}&=&F_{\rm \nu,max}~\nu_{\rm c}^{1/2}\nu_{\rm
m}^{\rm (p-1)/2}\nu_{\rm obs}^{\rm -p/2}
=4.0\times 10^{4}{\rm ergs~cm^{-2}}\nonumber\\
&& {\rm MeV^{-1}}~\epsilon_{\rm e,-1}^{\rm p-1}\epsilon_{\rm
B,-2}^{\rm p-2\over 4}E_{\rm 47}^{\rm {p+2\over 4}}C_{\rm p}^{\rm
(p-1)} ({d\over 150{\rm pc}})^{-2}
t_{\rm obs,1}^{\rm 2-3p\over 4}\nonumber\\
&& (1+Y)^{-1}({h\nu_{\rm obs} \over 20{\rm MeV}})^{\rm -(p-1)\over
2}. \label{Fobs1}
\end{eqnarray}
The value of $Y$ is in order of $1$ for typical parameters.

With typical parameters, the synchrotron radiation from forward
shock can reach the GeV range. Meanwhile, the
synchrotron-self-Compton radiation and inverse Compton radiation
will also contribute high-energy photons. For the chosen parameters,
these are much weaker than the synchrotron component. Therefore, we
ignore them. For typical parameters, the gamma-ray emission from a
pulsar outburst is described by a power-law spectrum with an index
of -1.1 \citep{Sari98}. From the X-ray and radio afterglow
observations of GW170817/GRB 170817, the spectral index is about
-0.6 \citep{Troja18,Ruan18}, which indicates that the cooling break
has not passed through the X-ray band. If the break passes through
the X-ray band, the spectral index becomes steeper to -1.1, which is
well consistent with the standard afterglow model.

Using the GEANT4 simulation code with QGSP-BERT-HP
, we computed the production yield of $^{14}$C due to gamma-rays in the
atmosphere. High-energy photons with energy from 1 MeV to 300 MeV
are considered. In order to calculate the energy required by the
increase of $^{14}$C, we apply GEANT4 code (version 10.5) to
simulate the production rate of $^{14}$C. This simulation code uses
Monte Carlo method to trace particles involved in the interactions.
The American standard atmosphere is used in our analysis.
Considering the plane-parallel atmosphere model, we construct an
atmosphere 87 km deep with 1 km resolution. The physics listed in
QGSP\_BERT\_HP are used to simulate the physics process. For the
gamma-ray photons, a power-law spectrum with the index $ \alpha =
-1.1 $ \citep{Sari98} is adopted with the minimum energy $ E_{\rm min}
= 1 \rm{MeV} $ and the maximum energy $ E_{\rm Max} = 100 \rm{MeV}
$. 3,000,000 photons satisfying the given spectrum are used in
simulation. From the simulation, we collect 6894 $^{14}$C atoms and
the average production rate of $^{14}$C is 60.67 atoms erg$^{-1}$.
For the AD 775 event, the required $^{14}$C in atmosphere is $ Q =
1.3 \times 10^8$ atoms cm $ ^{-2} $. The required energy is $
7\times 10^{24} $ erg in the atmosphere. The incident energy
necessary for the increase of $^{14}$C content in the atmosphere for
AD 775 event is about $E_{\rm atm}\sim 7\times10^{24}$ erg. Yields
of $^{10}$Be and $^{36}$Cl are about 0.36 and 0.12 atoms erg$^{-1}$,
respectively. The transport and deposition of $^{10}$Be would be
affected by gamma-ray radiation from outbursts \citep{Thomas05}, and
is not fully understood \citep{Heikkila09}. Hambaryan \& Neuh\"{a}user
(2013) found that the ratio between $^{14}$C and $^{10}$Be can be
explained in the gamma-ray photon case.

Below, we calculate the gamma-ray energy received in the atmosphere
from pulsar bursts. The fluence $F$ received in the atmosphere
between 1 MeV and 300 MeV from 1 second to $10^4$ seconds
\begin{equation}
F=\int_{1\rm MeV}^{300\rm MeV}\int_{1s}^{10^4s}f_{\nu_{\rm
obs}}dt_{\rm obs}d\nu_{\rm obs}
\end{equation} can be calculated numerically,
where $f_{\nu_{\rm obs}}$ is defined in Eq.(\ref{Fobs1}). The energy
received by the atmosphere can be calculated from $E_{\gamma}=\pi
R_{\oplus}^2\times F$ (with Earth radius $R_{\oplus}$). In addition,
the relativistic outflow from pulsar outbursts usually has a small
opening angle $\theta$ \citep{Yamazaki05,Lyutikov06}. For example,
Yamazaki et al. (2005) derived that the opening half-angle of the
jet of SGR 1806-20 giant outburst is 11$^\circ$ from light curve
\citep{Yamazaki05}. From $E_{\rm atm}=E_{\gamma}$, in order to
produce these $^{14}$C events, the distance of pulsar should be less
than $\sim 200$ pc, if $\theta\sim16^\circ$ is adopted.

There are four $^{14}$C events in about 5000 years. The rate of
$^{14}$C events is about 1000 years per event. Below, we estimate
the pulsar outburst rate. The characteristic size of multipole field
is $L\sim10^5$cm and hence the maximum number of multiple field
patches on the neutron star surface is $N\sim4\pi R^2/L^2\sim 10^3$.
It takes about $10^6$ year for the core field to diffuse out
\citep{Beloborodov16}. Therefore assuming each patch outburst
randomly connect over $10^6$ years, the mean outburst interval is
$10^6/10^3\sim 10^3$ years, which is consistent with the rate of
$^{14}$C events. Assuming there are about $10^4$ young pulsars in
the Milky Way galaxy, a typical rate of pulsar outbursts is ten per
year, which could be detectable by Fermi/GBM. Obviously, there is
lack of detection of such pulsar outbursts. The possible reason is
that the beaming effect of pulsar outbursts. Interestingly, a
handful of flares from non-SGR sources have been reported
\citep{Tavani11,Tam18} , but these amount to 3 over the past 25
years of high-energy observations.

Below, we compare the physical properties of the SGR 1806-20 giant
outburst with our proposed pulsar outburst. For the SGR 1806-20
giant outburst, the photon spectrum of the hard spike for giant
outburst is uncertain. It can be described by a blackbody spectrum
\citep{Hurley05} or a non-thermal spectrum \citep{Palmer05}. For
pulsar outbursts, we use the standard afterglow model of GRBs. The
high-energy photon spectrum is power-law form with index about -1.1
\citep{Sari98}. The isotropic gamma-ray energy of the SGR 1806-20
giant outburst is about $3.7\times 10^{46}$ erg in the spike. It
must be noted that the spike's intensity drove all X-ray and
gamma-ray detectors into saturation \citep{Palmer05,Hurley05}.
Therefore, the above gamma-ray energy may be a lower limit. The peak
gamma-ray luminosity of the SGR 1806-20 giant outburst is $2\times
10^{47}$ erg/s \citep{Hurley05}. From the radio afterglow fitting,
the total kinetic energy in the ejecta of SGR 1806-20 giant outburst
is about $10^{45}$ erg \citep{Taylor05}. The total energy of pulsar
outburst in our model is about $10^{47}$ erg (Eq.(1)). The rate of
giant outbursts is very uncertain. Lazzati et al. (2005) derived an
upper limit for the rate of giant outbursts: <1/130 yr$^{-1}$ per
galaxy. Popov \& Stern (2006) found that the rate of giant outburst
is less than $10^{-3} \rm yr^{-1}$ for Milky Way galaxy. In order to
explain the rate of $^{14}$C rapid-increase events, the rate of
pulsar outbursts is about $10^{-3} \rm yr^{-1}$. Yamazaki et al.
(2005) derived that the half opening angle of the jet of SGR 1806-20
giant outburst is 11$^\circ$ from the light curve. The half opening
angle of pulsar outburst used in this paper is 16$^\circ$.

\subsection{Transport and deposition of atmospheric $^{10}$Be}
Around AD 775 and AD 994, $^{10}$Be measurements from both Arctic
and Antarctic ice cores show possible excesses \citep{Mekhaldi15}.
However, the complexity of ice dating and poor temporal resolution
(about ten years) do not allow a reliable connection of the
measurements with the discussed $^{14}$C events. Meanwhile,
transport and deposition of atmospheric $^{10}$Be is very different
from $^{14}$C. For the atmospheric mixing of $^{10}$Be, it must be
noted that the pulsar outburst case is totally different from the
SPE case. Unlike protons, the high-energy photon emission of pulsar
bursts would deplete the ozone layer and affect atmospheric mixing
\citep{Thomas05,Melott17}. Besides the nuclide generation, radiation
from pulsar outburst must produce strong ionization of the upper
atmosphere and yield large amounts of nitric oxides in the
stratosphere. NOx are a catalyst of ozone depletion. The
stratospheric ozone is responsible for the temperature inversion
above the tropopause. Its concentration decline caused by burst
emission may reach 30\% at the equator and up to 70\% in the polar
regions \citep{Thomas05}. Ozone depletion will cause temperature
decrease in the lower stratosphere. Thus, the tropopause could rise
up and it could become more transparent, which would lead to
accelerated stratosphere-troposphere exchange. Stratospheric air
injections would give additional input of $^{10}$Be to the
troposphere \citep{Pavlov13}. Therefore, in this case, the
atmospheric mixing of $^{10}$Be would more complex than that of SPE
case, and is not well understood \citep{Heikkila13}.

\section{Spectrum of electrons observed near the Earth}
The electrons produced during the outbursts of pulsars can also
reach the Earth and affect spectrum of cosmic rays. However, due to
scattering on the interstellar turbulence it takes much longer for
electrons to arrive to Solar system as compared to gamma-rays.
Therefore, the spectrum observed in the Earth vicinity may be formed
by multiple eruptions. Given the propagation time is much longer
than the period between two successive eruptions one can ignore
variability of the sources and assume that they are almost
stationary. One should note however that this assumption is only
valid for calculations of the spectrum of particles far enough from
the source. In the close vicinity of the source its temporal
variability strongly affects the spectrum and spatial distribution
of particles.

There are two main irregularities in the spectrum of cosmic ray
electrons: the positron excess observed by PAMELA \citep{pamela} and
AMS-02 \citep{ams-pos} and spectral break in the spectrum observed
by AMS-02 \citep{ams}, CALET \citep{adr2018} and DAMPE
\citep{dampe}. We note that data from AMS-02 \citep{ams} and CALET
\citep{adr2018} are different from data obtained by DAMPE
\citep{dampe} as well as data obtained by Fermi-LAT \citep{fermi17}.
The reason for the discrepancy is unknown. In addition, recent
observations of the electron spectrum by VERITAS found no
irregularities \citep{veritas}, however their systematic
uncertainties are higher than the effect in question. Keeping this
in mind, we rely on data obtained by CALET and we estimate the
energy requirements necessary to make an observable contribution to
the spectrum of electrons.

These two features, namely the positron excess and the spectral
break, should not be confused. They share a similar concept
implicating that there is an additional local source of positrons
and electrons. But properties of sources required for positron
excess and are different from the ones required for spectral break
\citep{agu19}.

The positron excess was originally attributed to nearby pulsars
\citep[see, e.g.][and references therein]{hoop09, yuksel09}.
Therefore, below we are going to check if pulsars can also explain
the spectral break. Similarly to the papers mentioned above we take
Geminga as example, however we do not exclude that other pulsars can
contribute to the spectrum as well.

In the case of uniform background conditions the time-dependent
spectrum of particles arising from distant source can be calculated
from the diffusion equation
\begin{equation}\label{pr_state}
\frac{\partial f}{\partial t}  - \nabla \left(D\nabla f \right)+
\frac{\partial}{\partial E}\left( \frac{dE}{dt} f\right) = Q(E,{\bf
r},t)\,,
\end{equation}
where $dE/dt$ is the rate of energy losses, $D$ is the diffusion
coefficient and $Q$ describe spatial, temporal and spectral
properties of the source.

If medium around the source is uniform and source is small enough to
be considered as point-like, one can  use a spherically-symmetric
geometry. After introducing new variables
\begin{equation}
\tau(E,E_0)=
\int\limits^E_{E_0}{{dE}\over{dE/dt}}~~~~~\mbox{and}~~~~~\lambda=\int\limits^E_{E_0}{{D(E)}\over{dE/dt}}dE
\end{equation}
the Green function can be expressed as
\begin{equation}
G_k({\bf r},E,t;E_0,t_k)={{\mid dE/dt(E_0)\mid}\over{\mid
dE/dt(E)\mid}} {{\delta(t-t_0-\tau)}
\over{(4\pi\lambda)^{3/2}}}\exp\left[-\frac{{\bf
r}^2}{4\lambda}\right]
\end{equation}
and the solution of the equation is
\begin{equation}
f({\bf r},E,t)=\int\limits_0^t dt_0\int\limits_E^\infty dE_0
Q(E_0,t_0)G_k({\bf r},E,t;E_0, t_0) \label {sol11}
\end{equation}

If a stationary source ($\frac{\partial Q}{\partial t} = 0$) is
located close enough to the observer for energy losses to be
irrelevant ($r^2 \ll \lambda$)  but far enough for diffusion
approximation to be valid, spatial distribution of the particles can
be approximated by
\begin{equation}
f(r) \propto r^{-1} \,.
\end{equation}
In this case the differential flux of particles through a sphere of
radius $r_0$ is
\begin{equation}
S(E) = 4\pi r_0^2 D \frac{\partial f}{\partial r} = 4\pi Dr_0 f(r_0,
E) \,.
\end{equation}

There are two ways how to estimate the magnitude of $f$ and
therefore to calculate the contribution of the source into the
cosmic rays spectrum. The most natural way is to limit the total
power of source $Q$. However in the past the total energy output of
pulsar may have been significantly higher. Therefore the pulsar may
have been producing much more electrons in the past.

The second way, which takes into account past history, is to use
gamma-ray data. Recent observations by {\it HAWC} \citep{abe17} and
{\it Milagro} \citep{milagro} revealed an extended gamma-ray
emission around two pulsars, Geminga and B0656+14. This emission is
most likely leptonic and therefore its existence confirms that
pulsars may act like sources of electrons and positrons.

The spatial extent of the emission however is very narrow ($\approx
10-20$ pc) and is of order of mean free path of TeV electrons if
diffusion coefficient is of order of the intragalactic one. In order
to reproduce the observed halo it is necessary to confine electrons
and positrons within this distance, and confinement radius usually
significantly exceeds the mean free path. Therefore it is necessary
to assume that the value of diffusion coefficient in the
neighborhood of the pulsars is at least 100 times smaller than in
the Galaxy and that it is similar to that predicted for standard
Bohmian diffusion \citep{hoop17}.

The problem is that for $D \approx 10^{27}$ cm$^{2}/$s it requires
about 6 Myr for particles to cross 200 pc which significantly
exceeds the age of Geminga. To solve this problem one may assume
that there is an additional convective transport of
electrons \citep{hoop17} or that the area of low diffusion coefficient
is of finite size \citep{fang18}.

The two-diffusion model can be approximated in the following way.
Let's assume that the diffusion coefficient within radius $r_D$
around pulsar is $D_1$, while diffusion coefficient in ISM is $D_0$.
According to Abeysekara et al. (2017), $D_1/D_0 \approx 0.01$.
Assuming that energy losses are not essential one can obtain that
\begin{equation}
f(r) \propto \left\{ \begin{array}{ll}
f_0\frac{D_0}{D_1}\left(\frac{r_D}{r} - 1 + \frac{D_1}{D_0}\right)\,, & \mbox{for }r < r_d \\
f_0\frac{r_D}{r}\,. & \mbox{for }r > r_d
\end{array}
\right.
\end{equation}
On the other hand, column density of the gamma-ray emitting
electrons is
\begin{equation}
N_e = \int f(r) dl =
\frac{D_0}{D_1}f_0r_D\left\{\ln\left(\frac{r_{max}}{r_{min}}\right)
- 1 + \frac{D_1}{D_0}\right\} \,,
\end{equation}
where radius of confinement of electrons responsible for gamma-ray
emission observed by {\it HAWC} is $r_{max} \approx 10$ pc and
$r_{min}$ is of order of mean free path. Therefore $r_{min} \approx
3D_1/c \approx 0.03-0.1$ pc.

Surface brightness of the gamma-ray emission can be obtained by the
integration of the column density of the electrons with
inverse-Compton cross-section
\begin{equation}
I_\gamma = \frac{1}{4\pi}\int dE \int dE_{soft}N_e(E)
c\rho(E_{soft})\frac{d\sigma_{IC}(E,E_\gamma,E_{soft})}{dE_\gamma}
\,,
\end{equation}
where $E_{soft}$ and $\rho(E_{soft})$ is the energy and density of
soft photons respectively.

Assuming that spectrum of electrons is power law with index of $-2$
like we already did before, one can estimate from gamma-ray data
that the flux of 1 TeV electrons at the boundary $r = r_D$ is equal
to
\begin{equation}
E^3 j_0(1~\mbox{TeV},r_D) =
145~\mbox{m}^{-2}\mbox{sr}^{-1}\mbox{s}^{-1}\mbox{GeV}^{2}~ \cdot
\left(\frac{D_0/D_1}{100}\right)^{-1}\left(\frac{r_D}{10~\mbox{pc}}
\right)^{-1} \,.
\end{equation}
At the position of Earth the flux of electrons is $j_0(r_0)r_0
\approx j_0(r_D)r_D$ and therefore
\begin{equation}
E^3 j_0(1~\mbox{TeV}, r_0) \approx
9~\mbox{m}^{-2}\mbox{sr}^{-1}\mbox{s}^{-1}\mbox{GeV}^{2}~ \cdot
\left(\frac{D_0/D_1}{100}\right)^{-1}\left(\frac{r_0}{200~\mbox{pc}}
\right)^{-1} \,, \label{eq:geminga_max}
\end{equation}
which is lower than the value observed by CALET by factor of $3-4$.
One should note however that we completely ignored energy losses
therefore more accurate calculations of the electrons flux will most
likely produce lower values. Therefore Geminga alone cannot explain
spectral break observed by CALET. Yet, it potentially has power to
explain positron excess observed by AMS-02, but one need to assume
that it produces electrons and positrons in 1:1 ratio and that was
successfully demonstrated by \citet{hoop17} and \citet{fang18}.

One should notice that TeV gamma-ray emission require high-energy
electrons. However the elections produced during outburst have
relatively low energy (see Eq. (\ref{gamma_emax})) and therefore it
is unlikely that they will be visible. In this case we can restrict
the amount of electrons using total power of the source.

As discussed above, the spectrum of electrons from the relativistic
outflow is  power-law and is $\propto E^{-2}$. If we know at least
one point of differential spectrum $f(E_0,r_0)$ and energy losses of
electrons with energy $E_0$ are negligible, one can estimate total
power of the source as
\begin{equation}
W = 4\pi Dr_0 \int dE~ Ef(E_0,r_0) \left(\frac{E}{E_0}\right)^{-2} =
4\pi Dr_0E_0^2 f(E_0,r_0) \ln \left(\frac{E_{max}}{E_{min}}\right),
\end{equation}
where $D$ is the diffusion coefficient.

According to the data of CALET \citep{adr2018} there is a feature in
the spectrum of electrons with magnitude of about $30$
m$^{-2}$sr$^{-1}$s$^{-1}$GeV$^2$ at the energy of about $E_0 = 1.2$
TeV. We can obtain
\begin{equation}
W = 2.3\times 10^{34}~\mbox{erg/s}~\left(\frac{D}{3\times
10^{29}~\mbox{cm}^2/\mbox{s}}\right)
\left(\frac{r_0}{150~\mbox{pc}}\right),
\end{equation}
where we assumed that $E_{max} = 10^3E_{min}$. Note that in the case
of Geminga ($r_0\approx 200)$ pc total power required $\dot{W}
\approx 3\times 10^{34}$ erg/s is comparable to the total spin-down
power of the pulsar. Therefore Geminga can contribute only partially
to the observed spectral break. The positron excess on the
other hand requires less energy and can be completely supplied by
spin-down power alone. This point agrees with the conclusions we
made above and which were based on gamma-ray data.

The magnetic outbursts may produce up to $3\times 10^{35}$ erg/s
(assuming 10\% acceleration efficiency) which is more than enough to
reproduce the break. This conclusion is also in agreement with the
fact that positron excess and spectral break are of different
nature.

\begin{figure}
\includegraphics[width=\textwidth]{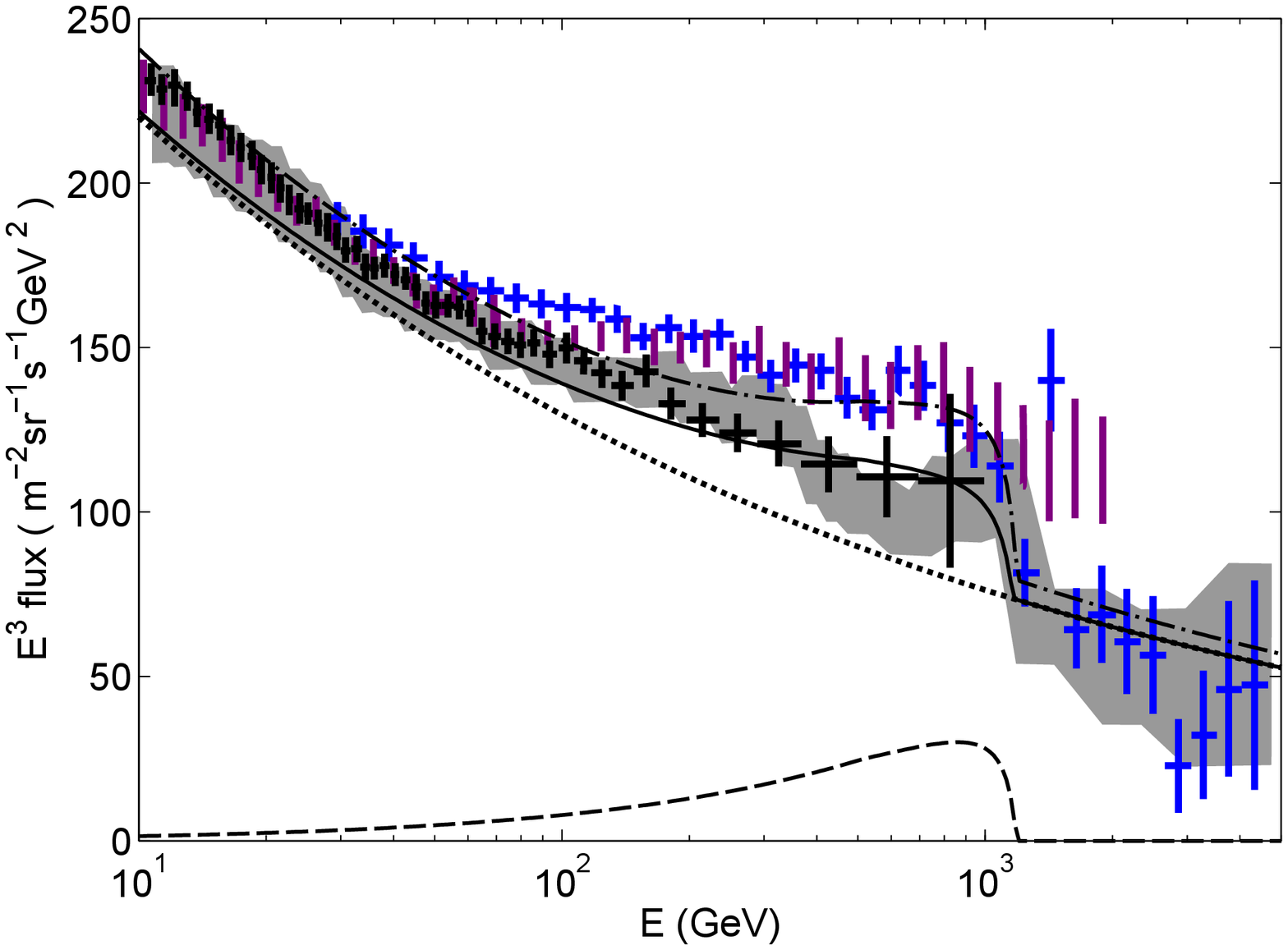}
\caption{Possible contribution of the nearby pulsars to the total
spectrum of electrons and positrons. Uncertainty band containing
quadratic sum of statistical and systematic errors for
CALET \citep{adr2018} data is shown by gray area. Data from
AMS-02 \citep{ams} are shown by black crosses, data from
Fermi-LAT \citep{fermi17} are shown by purple lines and data from
DAMPE \citep{dampe} are shown by blue crosses. The error bars ($\pm
1\sigma$) of AMS-02, Fermi-LAT and DAMPE include both systematic and
statistical uncertainties added in quadrature. Contribution of
pulsars necessary to fit CALET data is shown by a dashed line,
diffuse flux of the Galactic electrons is shown by dotted line,
total spectrum is shown by solid line. Fit to DAMPE data is shown by
dash-dotted line. For the latter we multiplied the diffuse flux by a
factor of 1.1 and contribution of the pulsars by a factor of 1.5.}
\label{fig:calet_spectrum}
\end{figure}

The expected contribution of the nearby pulsar to the total flux of
electrons and positrons obtained near the Earth is shown in Fig.
\ref{fig:calet_spectrum}. The spectrum was calculated using Eq.
(\ref{sol11}) and takes into account finite age of the pulsar (was
assumed of order of $4\times 10^5$ yrs) and energy losses.
Therefore, total power required to reproduce the observed
irregularity equals to $W = 3.6\times 10^{34}$ erg/s for distance of
$r_0 = 150$ pc.

If we rely on data obtained by DAMPE, we need slightly higher power
of electrons source. The possible fit to DAMPE data is shown in Fig.
\ref{fig:calet_spectrum} by dash-dotted line. The total power
required is $W = 5.5\times 10^{34}$ erg/s for distance of $r_0 =
150$ pc.

Since the source locates relatively close to the Earth, it should
produce a noticeable anisotropy in the arrival directions of
electrons. It also may be useful for differentiation between the
multi-burst model and the stationary model, since spectral shape in
both cases is effectively the same.

The dipole anisotropy in the arrival direction of electrons can be
estimated as
\begin{equation}
a = \frac{3D}{c} \frac{|\partial f / \partial r|}{f_{\rm tot}} \,,
\end{equation}
where $f_{\rm tot}$ is total spectrum of electrons. If the source is
stationary and energy losses are negligible, the expression can be
simplified to
\begin{equation}
a = \frac{3D}{cr_0} \frac{f_0(E, r_0)}{f_{\rm tot}(E, r_0)} \,.
\end{equation}
Therefore $a(E)$ can be easily estimated if we know energy
dependance of diffusion coefficient $D(E)$, total spectrum of
electrons $f_{\rm tot}(E, r_0)$ and contribution of local source
$f_0(E, r_0)$. In our calculations we assumed that $D = D_4
\left(\frac{E}{4~\rm{GeV}} \right)^{0.3}$ \citep{galprop}.

In the case of non-stationary source which injects electrons with a
series of bursts, solution of Eq. (\ref{sol11}) is required. In
addition to the injections responsible for the events happened AD
994, AD 775, BC 660 and BC 3371 we also assumed that source was
stationary until 6000 years ago to take into account all possible
injections happened in the past. Total power of each injection was
assumed to be $2\times 10^{45}$ erg.

\begin{figure}
\includegraphics[width=\textwidth]{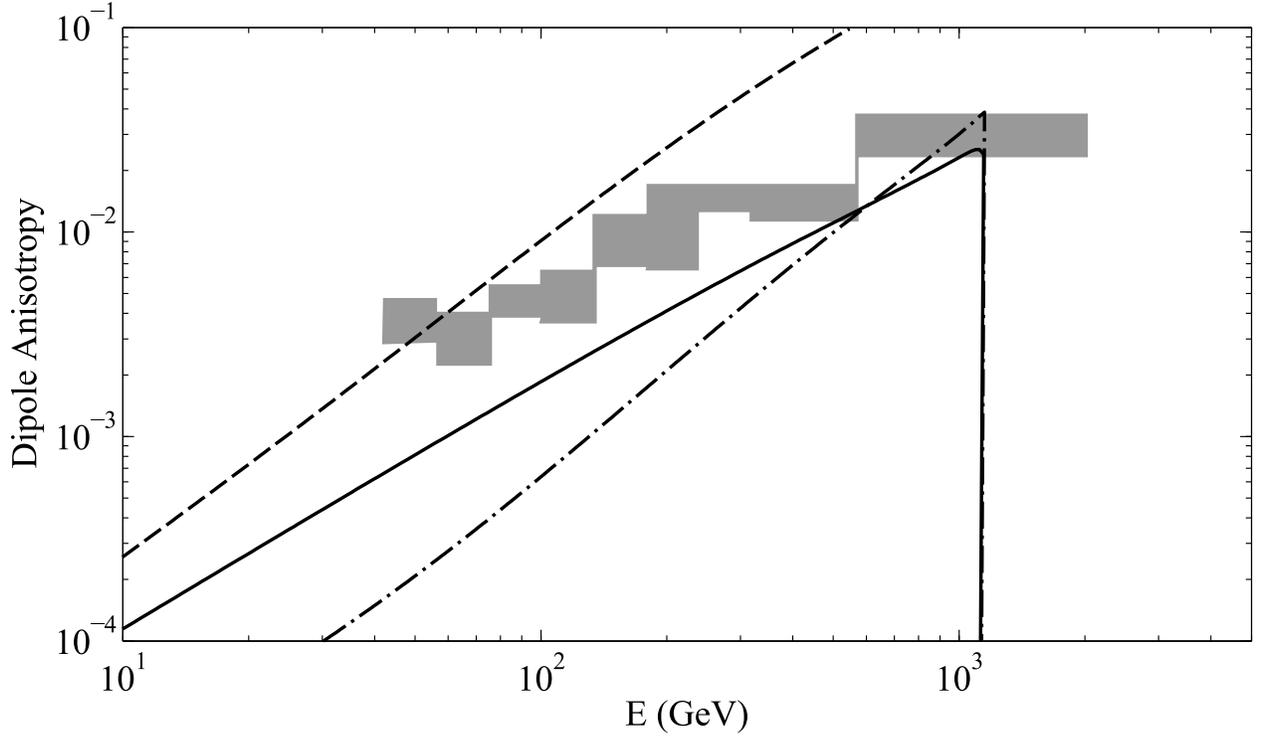}
\caption{Dipole anisotropy of electrons and positrons produced by
local source. Local diffusion coefficient is assumed to have a form
$D = D_4 \left(\frac{E}{4~\rm{GeV}} \right)^{0.3}$. The anisotropy
expected for stationary source is shown by solid line (with $D_4 =
5\times 10^{28}$ cm$^2$/s), while for the source with burst-like
injection it is shown by dashed line (with $D_4 = 5\times 10^{28}$
cm$^2$/s) and by dash-dotted line (with $D_4 = 10^{28}$ cm$^2$/s).
Gray bars show upper limits at 95\% confidence level according to
Fermi-LAT data \citep{fermi17a}.} \label{fig:anisotropy}
\end{figure}

The expected dipole anisotropy with experimental upper
limits obtained by Fermi-LAT \citep{fermi17a} is shown in Fig.
\ref{fig:anisotropy}. One can see that for the
burst-like source expected anisotropy is higher and also the
dependence on the energy is steeper. This difference may be the way
to differentiate the two given models.

\section{Neutron stars in our neighborhood}

We also search for the pulsars in our neighborhood. There are two
criteria to select candidates. First, the distance of pulsars must
be less than 200 pc. Second, due to the outburst, X-ray emissions of
pulsars should be detected as residual emission of crust heating. We
start by using the Australia Telescope National Facility Pulsar
Catalog (version 1.59) \citep{Manchester05} to compile a list of
candidates at distance $\leq200$ pc from the sun. Although, the
detection of a TeV gamma-ray halo around PSR B0656+14 provides good
reason to consider it a source of energetic electrons and positrons
\citep{abe17}. However, the distance of PSR B0656+14 is
$288^{+33}_{-27}$ pc \citep{Brisken03}. Therefore, we exclude it.
Among these candidates, we exclude the millisecond pulsars (i.e.
$P\leq20$~ms) as there is no evidence that outburst from these old
objects can attain the required level of energy release. For the
remaining candidates, if any of these sources has exhibited such
magnetar-like outburst, X-rays should be powered by crust heating.
With these selection criteria, we narrow down to four nearby X-ray
emitting pulsars as promising candidates, namely PSR~B1055-52,
PSR~B0834+06, Geminga, and RX~J1856.5-3754. Due to the pulsar
beaming effect, it must be noted that our census of  the young
nearby pulsars is not complete. It may be improved via HAWC TeV
gamma-ray survey observations in the near future, i.e., via further
TeV halo discoveries. Sudoh et al. (2019) have predicted that HAWC
can eventually detect up to $\sim80$ TeV halos. On the other hand,
with the upcoming Galactic plane survey with Cherenkov Telescope
Array, $\sim30-160$ new TeV halos can possibly be uncovered. Also,
it has been speculated that a few tens of currently unidentified
sources in TeV catalogs can possibly be TeV halos (Sudoh et al.
2019). From their observational properties, PSR~B0834+06, Geminga,
and RX~J1856.5-3754 could have had outbursts in the past.
Especially, the spin-down luminosity of the X-ray dim isolated
neutron star (XDINS) RX~J1856.5-3754 at $\sim108-134$~pc as inferred
from period $P$ and spin-down rate $\dot{P}$ is
$\dot{E}\sim3\times10^{30}$~erg/s, which is smaller, by an order of
magnitude, than its X-ray luminosity $L_{x}\sim3\times10^{31}$~erg/s
estimated at a distance of $d\sim120$ pc \citep{Burwitz03}. This
clearly indicates that the X-ray emission from RX~J1856.5-3754 is
not powered by its rotational energy. On the other hand, the energy
stored in its strong magnetic field is sufficient to power the
observed X-rays. Comparing the observed surface temperature of
RX~J1856.5-3754 with various neutron star cooling curves
\citep{Aguilera08}, it is far higher than the expected temperature
at its characteristic age even with the effect Joule heating taken
into account. Although the discrepancy can possible be reconciled if
its true age $4\times10^{5}$ years is closer to it kinematic
estimate \citep{Mignani17}, it may also indicate this neutron star
might have been heated by certain event, such as outburst, in the
past.

Recently, on top of the well-studied thermal radiation from a
two-temperature blackbody, an X-ray excess in keV regime has been
discovered from RX J1856.5-3754 \citep{Yoneyama17}. While the nature
of this excess remains to be uncertain, a possible scenario of
resonant cyclotron scattering has been proposed \citep{Yoneyama17}.
This suggests the additional component arises from the scattering
between the thermal surface photons and plasma flowing along the
close magnetic field lines, which have also been observed in the
X-ray spectra of magnetars \citep{Rea08}. If this explanation is
correct, it will further indicate the presence of magnetic
activities on RX J1856.5-3754 \citep{Beloborodov13}. X-ray
polarization measurement in the near future can provide a test for
this scenario \citep{Weisskopf18}.

PSR~B1055-52 is one of the earliest detected $\gamma-$ray pulsars
\citep{Fierro93}. Its dispersion measure suggests a distance of only
$d\sim90$~pc. It has a spin period of $P=0.197$~s and a spin-down
rate of $\dot{P}=5.8\times10^{-15}$~s/s. These imply a
characteristic age of $\tau\sim5\times10^{5}$~yrs and a dipolar
surface magnetic field strength of $B_{s}\sim1.1\times10^{12}$~G.
The X-ray emission of PSR~B1055-52 consist of two thermal component
and a non-thermal magnetospheric component \citep{De05}. The high
temperature ($T\sim2\times10^{6}$K) thermal component originates
from a hot spot with a radius of $\sim460$~m which is consistent
with the conventional polar cap size as defined by the footprints of
dipolar magnetic field lines on the surface. Hence, there is no hint
of the presence of stronger multipolar magnetic field on the
surface. We do not find any compelling feature from PSR~B1055-52
that makes it a possible contributor to the episodic cosmic ray
increment.

The distance of PSR~B0834+06 is estimated to be $d\sim190$~pc as
derived from its dispersion measure. Its spin parameters $P=1.27$~s
and $\dot{P}=6.8\times10^{-15}$~s/s imply a characteristic age of
$\tau\sim3\times10^{6}$~years and a dipolar surface magnetic field
strength of $B_{s}\sim3\times10^{12}$~G. The X-ray emission from
PSR~B0834+06 is found to be thermal dominant and can be described by
a hot spot on the stellar surface with a temperature of
$T\sim2\times10^{6}$~K \citep{Gil08}. The size of the hot spot is much
smaller than the polar cap defined by the dipolar field lines. This
suggests the surface field might be multipolar and could be $\sim50$
times stronger than the dipolar estimate given above. A field
strength of the order of $\sim10^{14}$~G makes outburst
activities possible in PSR~B0834+06, though there is so far no
observational indication for any transient behavior from this
pulsar.

Geminga (PSR~J0633+1746) is the second strongest persistent
$\gamma-$ray source in the sky. The distance of this radio-quiet
$\gamma-$ray pulsar is in the range of $d\sim190-370$~pc as
determined by the parallax measurement of its optical
counterpart \citep{Faherty07}. Its spin period and the proper-motion
corrected spin-down rate ($P=0.237$~s,
$\dot{P}=1.1\times10^{-14}$~s/s) suggest a characteristic age of
$\tau\sim3\times10^{5}$~yrs and a dipolar surface field strength of
$B_{s}\sim1.6\times10^{12}$~G. X-ray observations have identified
small hot spot on its surface \citep{Caraveo04}. The area of this hot
spot is $\sim25$ times smaller than the dipolar polar cap. Although
the discrepancy can be due to a geometrical viewing effect, it can
also possibly be a result of multipolar surface field. In such
scenario, the surface field strength can attain a value of
$\sim4\times10^{13}$~G which is comparable to the field of low
magnetic field magnetars \citep{Turolla13}. Some previous studies of
the distribution of thermal emission on the surface of Geminga have
even suggested the thermal X-rays can possibly be a ``mark" left by
tectonic activity in the past \citep{Page95,Halpern93}.

Geminga is also powering a complex pulsar wind nebula (PWN) which
can be resolved in several spatial component using X-ray
\citep{Hui17,Posselt17}. In a multi-epoch analysis, Hui et al.
(2017) have discovered fast X-ray variabilities in various
components of Geminga PWN. In the region of the post-shock flow (cf.
Fig.~2 in \citep{Hui17}, clumps of X-rays have been found to appear
in several epochs. This resembles the magnetic reconnection in the
sun that gives rise to the eruptive coronal mass ejection (cf.
Fig.~1 in \citep{Lin05}. Energetic charged particles can possibly be
ejected by Geminga PWN in a similar way.

Different from the above three candidates, RX~J1856.5-3754 is not a
rotation-powered pulsar. It belongs to a class of X-ray dim isolated
neutron stars (see \citep{Haberl13} for a detailed review). Its
distance has been bracketed in a range of $\sim108-134$~pc
\citep{Mignani17}. The X-ray emission from RX~J1856.5-3754 is purely
thermal \citep{Sartore12}. While the thermal X-rays of the hot spot
from the rotation-powered pulsars likely comes from the bombardment
of the back-flow current from the accelerating regions in their
magnetosphere \citep{Cheng99}, the thermal radiation from
RX~J1856.5-3754 should have a different origin. A low amplitude
X-ray pulsation at a period of $P=7.06$~s \citep{Tiengo07} and its
subsequent spin-down rate $\dot{P}=3.0\times10^{-14}$~s/s
\citep{van08} have been found. These imply a surface dipolar field
strength of $B_{s}\sim1.5\times10^{13}$~G and a characteristic age
of $\tau\sim4\times10^{6}$~years. Recently, optical polarization
measurements have confirmed the presence of strong magnetic field on
RX~J1856.5-3754 \citep{Mignani17}. It spin-down luminosity as
inferred from $P$ and $\dot{P}$ is
$\dot{E}\sim3\times10^{30}$~erg/s, which is smaller than its X-ray
luminosity $L_{x}\sim3\times10^{31}$~erg/s \citep{Burwitz03} by an
order of magnitude. On the other hand, the energy stored in its
strong magnetic field is sufficient to power the observed X-rays.
Comparing the observed surface temperature of RX~J1856.5-3754 with
various neutron star cooling curves \citep{Aguilera08}, it is far
higher than the expected temperature at its characteristic age even
with the effect Joule heating taken into account. Although the
discrepancy can possible be reconciled if its true age is closer to
it kinematic estimate ($4\times10^{5}$~yrs) \citep{Mignani17}, it
may also indicate this neutron star may have been heated by certain
event, such as outburst, in the past. The heat can come from
internal heating by plastic flow \citep{Li16} or external heating by
Alfv\'{e}n waves launched in the magnetosphere \citep{Li15}. Such
heating is transient and produce delayed thermal afterglow. For its
impact on the long term cooling of the neutron star, we can estimate
the increased thermal luminosity by
\begin{equation}
\mathcal{L} \sim E/t_{\rm amb} \sim 10^{32} L_5^{13/5}\delta
B_{16}^{8/5} B_{13}^{8/5} (\rho/\rho_{\rm nuc})^{6/5} \rm erg/s.
\end{equation}

The currently known TeV halos around pulsars are resulted from the
interactions between the continuous outflow of electrons/positrons
with the ambient photon field. While the case that we consider here
was resulted from the episodic injection by RX J1856.5-3754 in the
past, the diffusion can make the flux of its possible relic halo
below the detection limit of the existing TeV observing facilities.

\section{Conclusions}
In this paper, we propose that the outbursts of nearby pulsars
($d<200$pc) can cause the $^{14}$C events and contribute a
significant fraction of cosmic electron spectrum in trillion
electron volts energy range. The relativistic outflow producing by
outbursts encounters the ISM, the external shock is generated,
including forward and reverse shocks, which can accelerate
electrons. The synchrotron emission from accelerated electrons
generates high-energy photons, which interact with the atmosphere
producing the cosmogenic nuclide $^{14}$C.

Meanwhile, these same relativistic electrons contribute a
significant fraction of the cosmic electron spectrum in the trillion
electron volts energy range. They may be responsible for the
spectral break recently observed in the spectrum. We also review the
properties of cataloged pulsars in the solar neighborhood, and find
that PSR B0834+06, Geminga, and RX J1856.5-3754 could have had
outburst activities in the past.

Our model can be tested with future experimental facts and
astronomical observations. First, the short time and the narrow beam
of the outburst would have an influence on the time and intensity of
the $^{14}$C increase at different locations on Earth. Second, our
model predicts a hard energy dependence of dipole anisotropy in the
arrival direction of electrons. Third, a non-thermal component is
expected in soft X-rays band. Since these outburst events can
significantly affect our environment, monitoring the nearby pulsar
outbursts are important in future.

\textbf{Acknowledgements} We thank the referee for detailed and very
constructive suggestions that have allowed us to improve our
manuscript. We also thank Xiang Li and Lei Feng for helpful
discussion on GEANT4 code, and Prof. Kevin Mackeown for language
editing. FYW is supported by the National Natural Science Foundation
of China (grant U1831207). DOC is supported in parts by the grant
RFBR 18-02-00075 and by foundation for the advancement of
theoretical physics and mathematics ``BASIS". CYH is supported by
the National Research Foundation of Korea grant 2016R1A5A1013277.
KSC is supported by a GRF grant under 17302315.

\end{document}